\documentclass[12pt]{article}
\usepackage{pdproc,epsfig,subfigure}

\begin{document}

\renewcommand{\thefootnote}{\alph{footnote}}
  
\title{
FROM AMANDA TO ICECUBE\\}

\author{PER OLOF HULTH FOR THE ICECUBE COLLABORATION\footnote{see http://icecube.wisc.edu/pub\_and\_doc/conferences/NOVE-06 for full list of authors}}

\address{ Department of Physics, Stockholm University, AlbaNova University Centre
 \\
  SE-106 91, Stockholm,Sweden\\
 {\rm E-mail: hulth@physto.se}}

\abstract { The success of the AMANDA neutrino telescope has shown that the ice sheet at the geographical South Pole is a suitable medium for optical Cherenkov detection of high energy neutrino interactions. Several thousands of atmospheric neutrinos have been recorded by AMANDA and the sensitivity for cosmic neutrinos has continuously improved. So far no cosmic neutrino signals have been detected. The deployment of the much larger  and more sensitive IceCube neutrino observatory has started and nine out of 80 strings have been installed. This paper summarizes some of  the results obtained by the AMANDA telescope and presents the status of the IceCube project.}

\normalsize\baselineskip=15pt

\section{Introduction}
A new window to the Universe is opened by using cosmic neutrinos as probes. Neutrinos traverse huge amounts of matter without being absorbed. They are not deflected by magnetic fields and thus point back to the source (e.g. the sources of the high energy cosmic rays). Only neutrinos might reveal information about hidden processes close to the source. 

Using different models\cite{wb,mpr} and the observed flux of high energy cosmic rays and  cosmic photons, the expected flux of cosmic neutrinos can be estimated.  This  results in an upper bound for the total diffuse flux of neutrinos  of $dN/dE_\nu \sim 5\cdot 10^{-8} E^{-2}_\nu$ GeV$^{-1}$ cm$^{-2}$ s$^{-1}$ sr$^{-1}$, demanding very large detector mass in order to get a significant signal. 

Detecting high energy cosmic neutrinos by using large volumes of water (or later ice)  as detector medium was proposed more than 40 years ago\cite {markov}. Today there are three\footnote{The ANTARES collaboration has since the conference deployed their first string} detectors in operation:  the Baikal telescope NT-200+ in the lake Baikal, the AMANDA telescope at the geographical South Pole and the first part of the new IceCube neutrino observatory at the same place. 

The only cosmic neutrinos detected so far are low energy  neutrinos from the fusion process in the Sun ($<20$ MeV) and a short burst in 1987 from the supernova SN1987a ($<$ 50 MeV) in the Large Magellanic Cloud. So far, no high energy cosmic neutrinos have been detected.
 
This paper describes the AMANDA neutrino telescope and some recent  results as well as  the status of the construction of the new IceCube Neutrino Observatory.

\subsection{Detection principle}
AMANDA and IceCube are detecting the emitted Cherenkov light from neutrino interactions in the optically transparent ice at the South Pole. High energy $\nu_{\mu}$ produce muons with a range in ice of several km (about 1 km at 200 GeV) allowing muons created far outside the instrumented detector volume to be detected. The mean angle  between the incoming neutrino and the out-going muon falls approximately as $E^{-0.5}$ and is about 1 degree at 1 TeV.

Charged current $\nu_e$ and  $\nu_{\tau}$ (at moderate energies) and neutral current interactions  will produce ``cascades'' in which most of the secondary particles will interact and stop within a few tens of metres. To first order the Cherenkov light will  look like it is coming from a point source inside the large detector volume. For  $\nu_{\tau}$ with energies above several PeV the decay length of the resulting tau will be hundreds of meters, allowing detection of two ``cascades" (``double bang events"), one from the primary interaction and a second from the decay of the tau.

\section{AMANDA}
\begin{figure}[ht]
\subfigure{
    \mbox{\epsfig{file=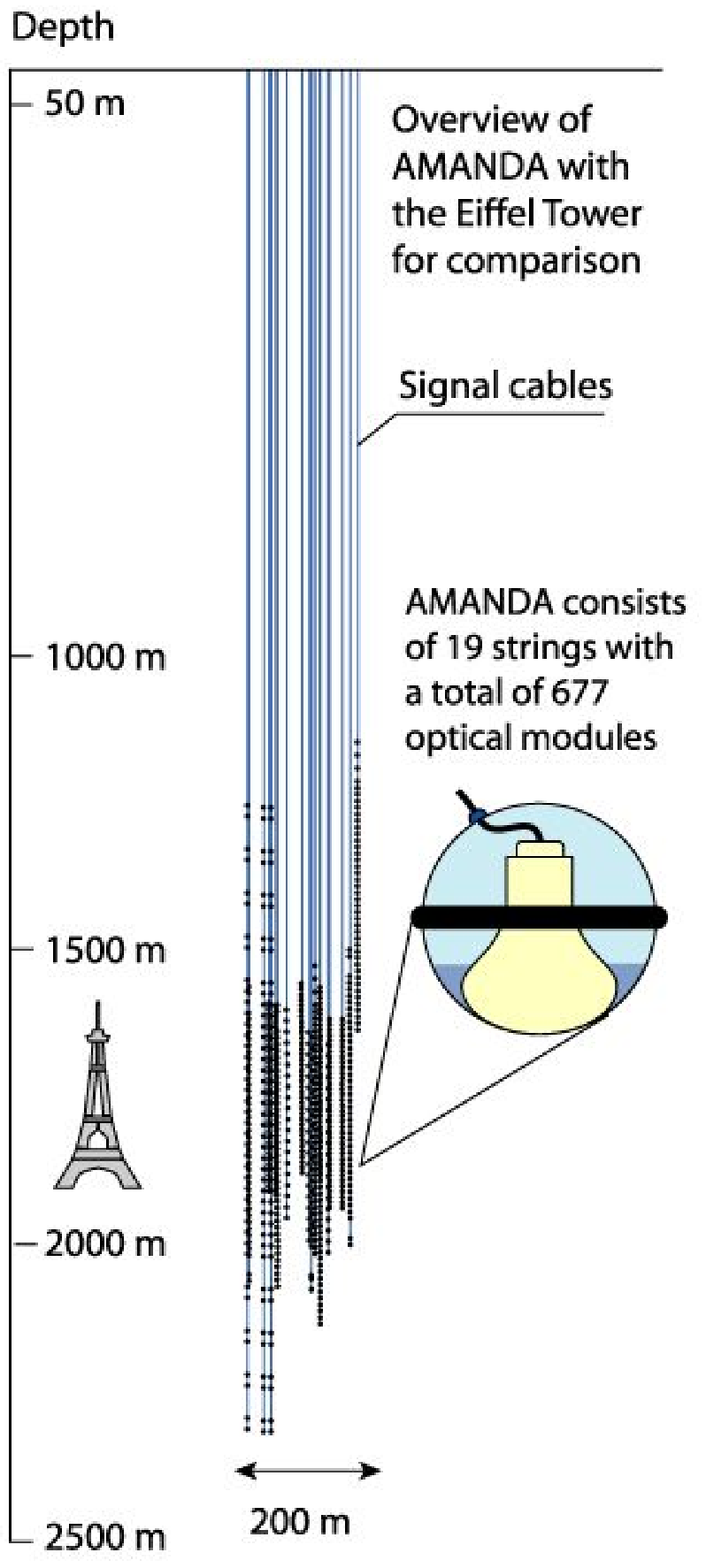,width=0.30\textwidth}}
}
\subfigure{
    \mbox{\epsfig{file=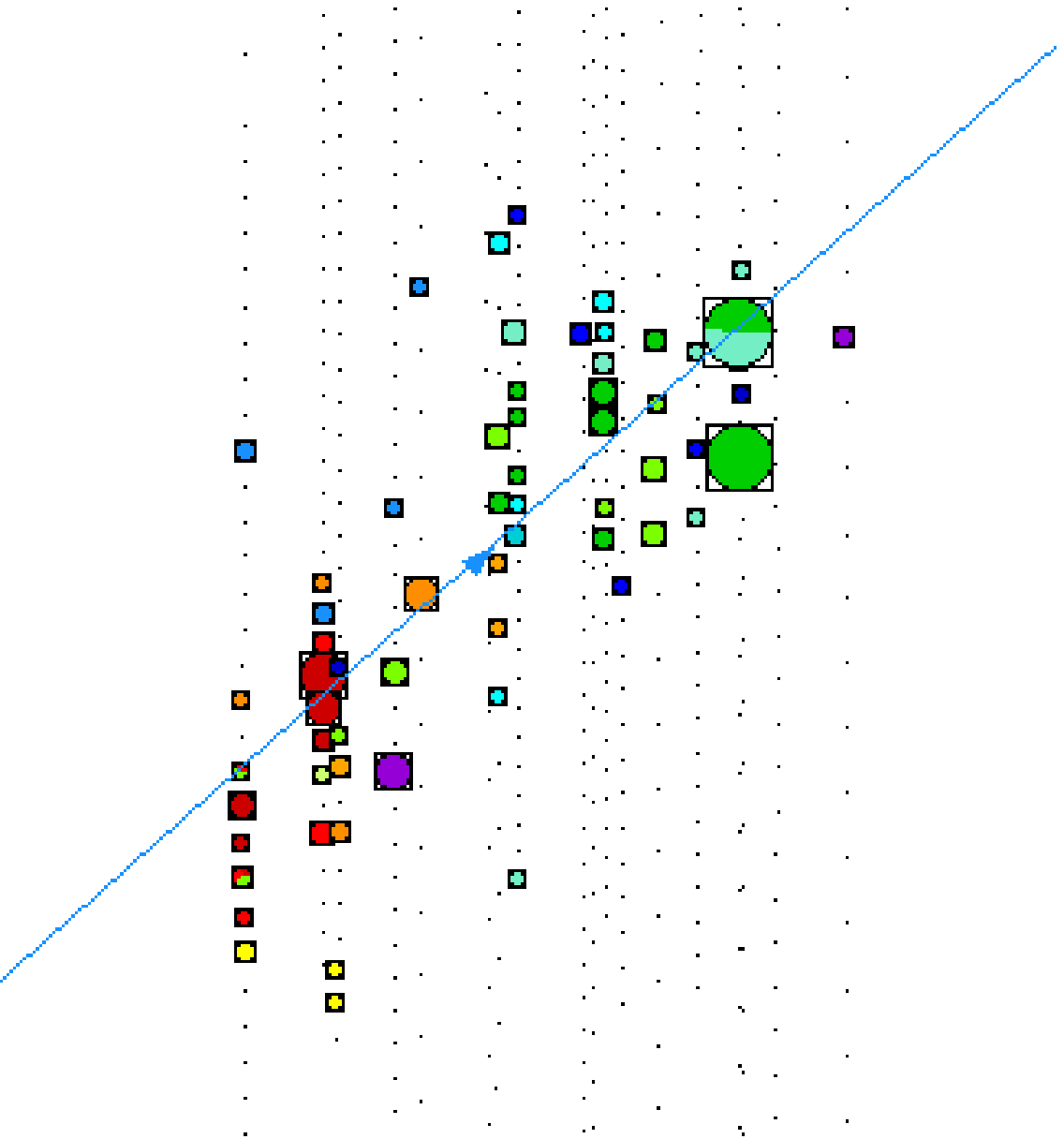,width=0.45\textwidth}}
}
\caption[2]{ (Left) The AMANDA Detector. (right) a recorded upward going muon event. The size of the squares corresponds to recorded amplitude}
\label{fig:amanda}
\end{figure}

The AMANDA telescope  (figure~\ref{fig:amanda})  consists of 19 strings with a total of 677 optical modules. The distance between the optical modules in a string varies between 10~m and 20~m. The main sensitive volume is situated between 1500 m and 2000 m below the snow surface. The effective area for AMANDA is about $10^{4}$~m$^{2}$  for 1 TeV muons. The 
optical modules are deployed in holes (about 60 cm in diameter) in the ice, which have been melted by a hot water drilling 
system. The telescope was completed in January 2000 and is continuously taking data except for short interruptions between November and February due to service.  Data were already taken by the partially finished AMANDA telescope using the first four and later 10 strings (AMANDA-B4 and AMANDA-B10).  AMANDA is running in coincidence with the air shower detector, SPASE-II situated on the surface. SPASE-II is observing the electron component of the air shower whilst AMANDA detects the muon component. The muons observed inside AMANDA from coincident events with SPASE are used for calibration of AMANDA as well as for composition studies of the cosmic rays.  

AMANDA is taking data with the completed detector for the seventh year. The AMANDA collaboration has published many results in different areas. A short summary of the most recent results is given below.

\subsection{Atmospheric Neutrinos}
The  cosmic rays interacting in the atmosphere produce a flux of atmospheric neutrinos. These are a background when searching for high energy cosmic neutrinos. The energy dependence of the atmospheric neutrino flux is about $E_{\nu_{\mu}}^{-3.7}$ above 1 TeV and differs from the expected  $E_{\nu_{\mu}}^{-2}$ for extra-terrestrial neutrinos. The atmospheric neutrinos can be used to study the efficiency of the detector. AMANDA has reported 3329 atmospheric neutrinos from the period 2000-2003.

AMANDA has used a regularised unfolding method to estimate the atmospheric neutrino energies from the observed upward going muon energies.   Figure~\ref{fig:atmospheric} (left) shows the  preliminary atmospheric neutrino flux for AMANDA data taken during 2000\cite{icrc-klas}. The maximum observed neutrino energy  is about 300 TeV (about 1000 times higher than available neutrino beams at FNAL or  CERN). The lower energy results from Frejus\cite{frejus} are also shown. The dotted curves in figure~\ref{fig:atmospheric} are the horizontal and vertical fluxes parameterized according to Volkova above 100 GeV and Honda below 100 GeV\cite{volkova}.

\subsection{Diffuse fluxes}

\begin{figure}
  \includegraphics[width=.47\textwidth]{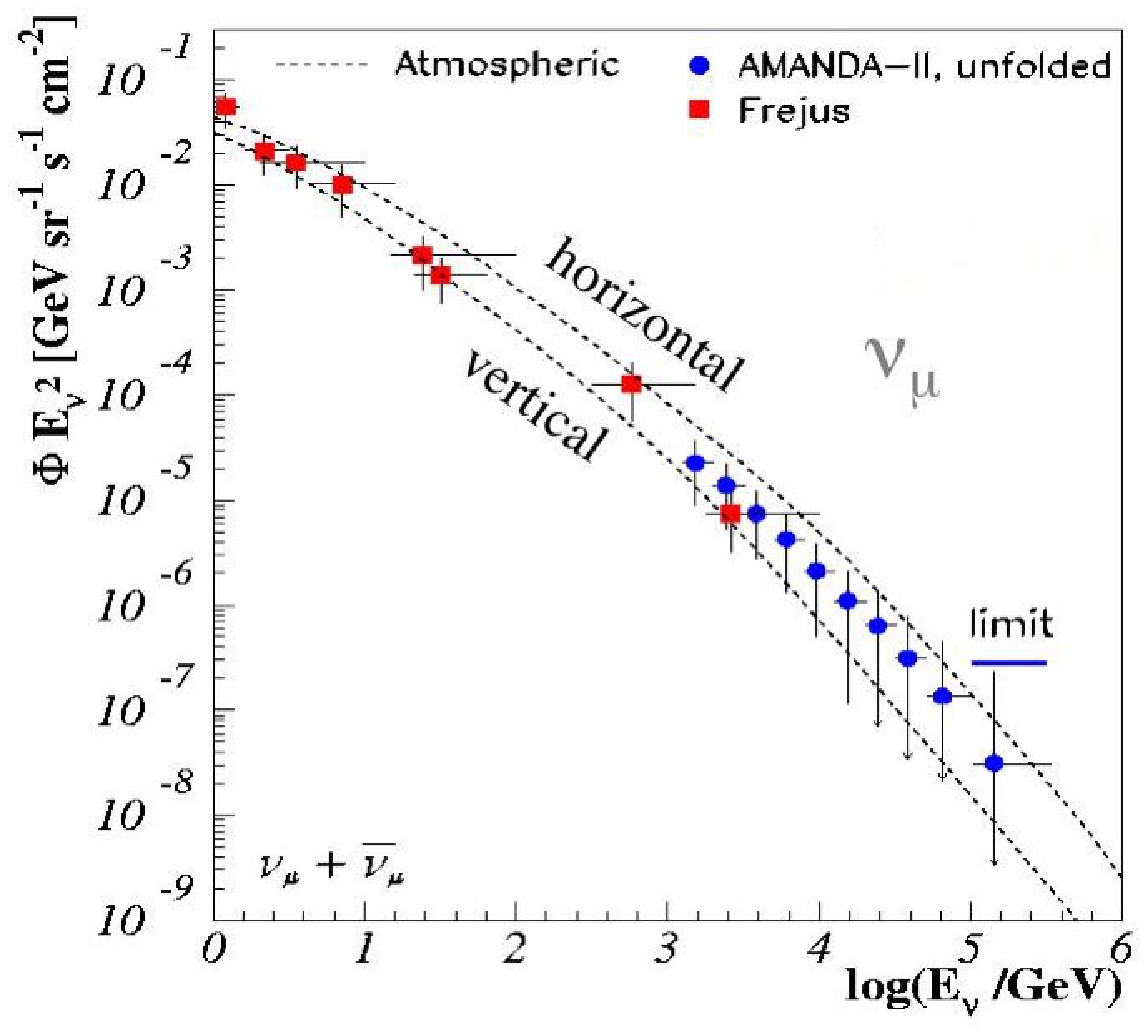}
  \includegraphics[width=.47\textwidth]{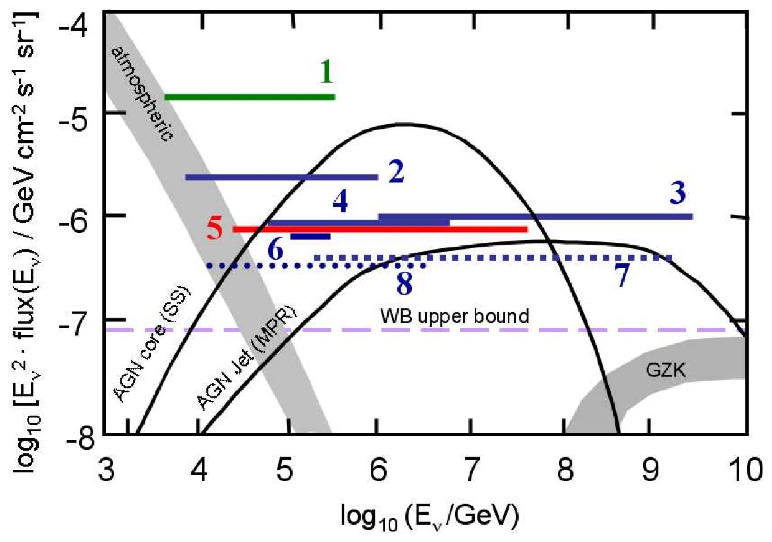} 
\caption[]{Left: Atmospheric neutrino flux vs. neutrino energy from AMANDA and the Frejus experiment \cite{frejus} . Right: All-flavour neutrino limits and sensitivity on an  ${E^{2}}$${dN\over dE}$ plot assuming neutrino oscillation and 1:1:1 flavor mixing. Neutrino oscillations are taken into account. (1) The MACRO  $ \nu_{\mu}$ analysis for 5.8 years (times 3)\cite{macro}. 
(2) The AMANDA-B10 $ \nu_{\mu}$ analysis from 1997 (times 3\cite{a97}.
(3) AMANDA-B10 ultra-high energy  neutrinos of all flavours 1997\cite{ua97}.
(4) AMANDA all-flavour cascade limit from 2000\cite{acascade}. 
(5) Baikal cascades 1998-2003\cite{bcascade}. 
(6) The preliminary results of the 2000 AMANDA $ \nu_{\mu}$ analysis\cite{icrc-klas} (times 3). 
This limit was derived after unfolding the upward going muon energies.
(7) The sensitivity for AMANDA  ultra-high energy neutrinos of all flavours 2000\cite{lisa}. 
(8) 2000 to  2003 AMANDA $ \nu_{\mu}$ sensitivity   (times 3)\cite{hodges}.
}
\label{fig:atmospheric}
\end{figure}

Neutrinos from different cosmic sources  with too few events to be detected individually will add up to a  ``diffuse" flux of cosmic neutrinos. Due to the expected harder energy spectrum the signal for diffuse cosmic neutrinos will be seen as high energy events above the atmospheric neutrino energy distribution. Figure~\ref{fig:atmospheric} (left) shows a 90 \% CL upper limit of  
$E^{2}$ $\Phi_{\nu_{\mu}}<2.6\cdot 10^{-7}$ GeV cm$^{-2}$ sr$^{-1}$ s$^{-1}$ for an  extraterrestrial $E_{\nu_{\mu}}^{-2}$ flux component  between 100 TeV and 300 TeV (line labeled 6 on the right).  To the right  several different limits for diffuse cosmic neutrinos (sum of all flavours) are shown. These are based on $\nu_{\mu}$ as well as cascade event ($ \nu_e$  and $ \nu_{\tau}$) analyses. The single neutrino flavour limits have been multiplied by a factor of three for neutrino oscillations to obtain the all flavour flux at Earth. The lines labeled 7 and 8 in figure~\ref{fig:atmospheric} (right) are not limits but expected sensitivities since the data are still blinded.  

The sensitivity for detecting a diffuse cosmic neutrino flux has improved by an order of magnitude in less than 10 years.

\subsection{Point source searches}

\begin{figure}[h]
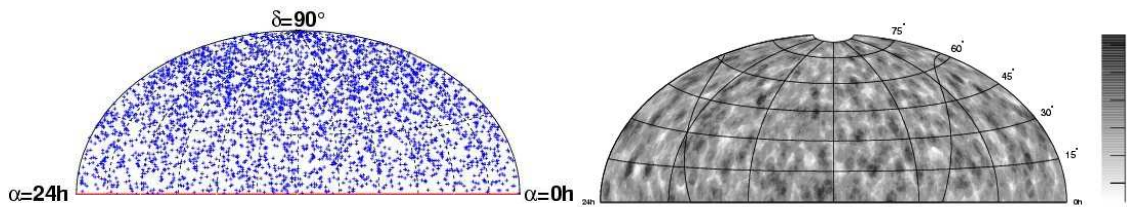

  \includegraphics[height=.12\textheight]{Bernardini_Fig1.epsi}
  \includegraphics[height=.11\textheight]{Bernardini_Fig2.epsi}
  \caption[]{Left: sky map of the selected 3329 neutrino events collected during 2000-03. Right: significance map
           for the search for clusters of events in the northern hemisphere, based on the selected events (gray
           scale is in sigma).}
  \label{fig:sky}
\end{figure}

The AMANDA collaboration has performed the most sensitive search for neutrino point sources in the northern sky using 3329 ${\nu_{\mu}}$  events (purity about 95 {\%}) obtained during the years  2000-2003\cite{acker}. Figure~\ref{fig:sky}  shows the directions in the northern sky (in celestial coordinates) and the corresponding  significance map for the AMANDA events. No evidence for any extraterrestrial point source is found. The highest excess corresponds to a significance of about 3.4${\sigma}$. The probability to observe this or a higher excess, taking into account  trial factors, is 92 \%.

A search for  33  preselected neutrino candidate sources including galactic and extra-galactic sources has been done  based on four years of data \cite{acker}.  The result is shown in Table~\ref{ps-table}. The highest excess is found in the direction of the Crab Nebula, with 10 observed events compared to an average of 5.4 expected background events (about 1.7${\sigma}$).  Taking into account the trial factor the probability that a background fluctuation produces this or a larger deviation in any of the 33 search bins is 64~\%, (due to the multiplicity of the directions examined and the correlation between overlapping search bins). 

\setlength\tabcolsep{0.18cm}
\begin{table*}[t]
\begin{minipage}{\textwidth}
\begin{center}
\begin{footnotesize}
 \begin{tabular}{lccccc|lccccc}\hline
\multicolumn{12}{c}{\bf Amanda search for neutrinos} \\
\multicolumn{12}{c}{\bf from 33 preselected objects} \\ \hline
     Candidate & $\delta$($^\circ$) & $\alpha$(h)    &
               $\!\!\!n_{\mathrm{obs}}\!\!\!$      & $\!\!\!n_{\mathrm{b}}\!\!\!$ &
               $\Phi_{\nu}^{\mathrm{lim}}$           
   &  Candidate & $\delta$($^\circ$) & $\alpha$(h)    &
               $\!\!\!n_{\mathrm{obs}}\!\!\!$      & $\!\!\!n_{\mathrm{b}}\!\!\!$ &
               $\Phi_{\nu}^{\mathrm{lim}}$           
\\\hline
\multicolumn{5}{c}{ \emph{TeV Blazars} } &  &  \multicolumn{5}{c}{ \emph{SNR \& Pulsars} }\\
     Markarian 421  & 38.2 & 11.07 & 6 & 5.6 & 0.68 &   SGR 1900+14    &  9.3 & 19.12 & 3 & 4.3 & 0.35\\
     Markarian 501  & 39.8 & 16.90 & 5 & 5.0 & 0.61 &   Geminga        & 17.9 &  6.57 & 3 & 5.2 & 0.29\\
     1ES 1426+428   & 42.7 & 14.48 & 4 & 4.3 & 0.54 &   Crab Nebula    & 22.0 &  5.58 &10 & 5.4 & 1.3\\ 
     1ES 2344+514   & 51.7 & 23.78 & 3 & 4.9 & 0.38 &   Cassiopeia A   & 58.8 & 23.39 & 4 & 4.6 & 0.57\\
     1ES 1959+650   & 65.1 & 20.00 & 5 & 3.7 & 1.0  &   &   &  &  &  &  \\
\multicolumn{6}{c}{ \emph{GeV Blazars} } & \multicolumn{6}{c}{ \emph{Microquasars} } \\
     QSO 0528+134   & 13.4 &  5.52 & 4 & 5.0 & 0.39 &  SS433          &  5.0 & 19.20 & 2 & 4.5 & 0.21\\
     QSO 0235+164   & 16.6 &  2.62 & 6 & 5.0 & 0.70 &  GRS 1915+105   & 10.9 & 19.25 & 6 & 4.8 & 0.71\\
     QSO 1611+343   & 34.4 & 16.24 & 5 & 5.2 & 0.56 &  GRO J0422+32   & 32.9 &  4.36 & 5 & 5.1 & 0.59\\
     QSO 1633+382   & 38.2 & 16.59 & 4 & 5.6 & 0.37 &  Cygnus X1      & 35.2 & 19.97 & 4 & 5.2 & 0.40\\
     QSO 0219+428   & 42.9 &  2.38 & 4 & 4.3 & 0.54 &  LS I +61 303   & 61.2 &  2.68 & 3 & 3.7 & 0.60\\
     QSO 0954+556   & 55.0 &  9.87 & 2 & 5.2 & 0.22 &  Cygnus X3      & 41.0 & 20.54 & 6 & 5.0 & 0.77\\
     QSO 0954+556   & 55.0 &  9.87 & 2 & 5.2 & 0.22 &  XTE J1118+480  & 48.0 & 11.30 & 2 & 5.4 & 0.20\\
		 &  & 	 &  & 	 &  &		       CI Cam         & 56.0 &  4.33 & 5 & 5.1 & 0.66\\
 
 \multicolumn{12}{c}{ \emph{Miscellaneous} }\\
     3EG J0450+1105 & 11.4 &  4.82 & 6 & 4.7 & 0.72 &  J2032+4131     & 41.5 & 20.54 & 6 & 5.3 & 0.74\\
     M 87           & 12.4 & 12.51 & 4 & 4.9 & 0.39 & NGC 1275       & 41.5 &  3.33 & 4 & 5.3 & 0.41\\
     UHE CR Doublet & 20.4 &  1.28 & 3 & 5.1 & 0.30 & UHE CR Triplet & 56.9 & 11.32 & 6 & 4.7 & 0.95\\
     AO 0535+26     & 26.3 &  5.65 & 5 & 5.0 & 0.57 & PSR J0205+6449 & 64.8 &  2.09 & 1 & 3.7 & 0.24\\
     PSR 1951+32    & 32.9 & 19.88 & 2 & 5.1 & 0.21 & & & & & \\ 
      \hline\hline
 \end{tabular}
\end{footnotesize}
 \caption{\label{ps-table}
               Results from the AMANDA search for neutrinos from selected objects.
               $\delta$ is the declination in degrees, $\alpha$ the
               right ascension in hours,
               $n_{obs}$ is the number of observed events, and $n_{b}$
               the expected
               background. $\Phi_{\nu}^{\mathrm{lim}}$ is the
               90\% CL upper limit in units of $10^{-8}
               \mathrm{cm}^{-2}\mathrm{s}^{-1}$
               for a spectral index of 2
               and integrated above 10 GeV.
               These results are preliminary (systematic errors
               are not included).}
\end{center}
\end{minipage}
\end{table*}
\setlength\tabbingsep{\labelsep}

\subsection{Gamma-ray bursts}
Gamma-ray bursts are candidate sources for the ultra high energy cosmic rays. They are the most energetic objects in the universe and observing neutrinos in coincidence with the bursts would give evidence for the acceleration of hadrons. Due to the coincidence in  direction and time with a GRB  such a search is almost free of background from atmospheric neutrinos. The AMANDA collaboration has searched for neutrinos from more than 500  GRBs without any observed signal, and derived the most stringent limits\cite{kuehn,stamatikos,hughey}.

\subsection{Searches for neutralino dark matter}

\begin{figure}
  \includegraphics[width=.47\textwidth]{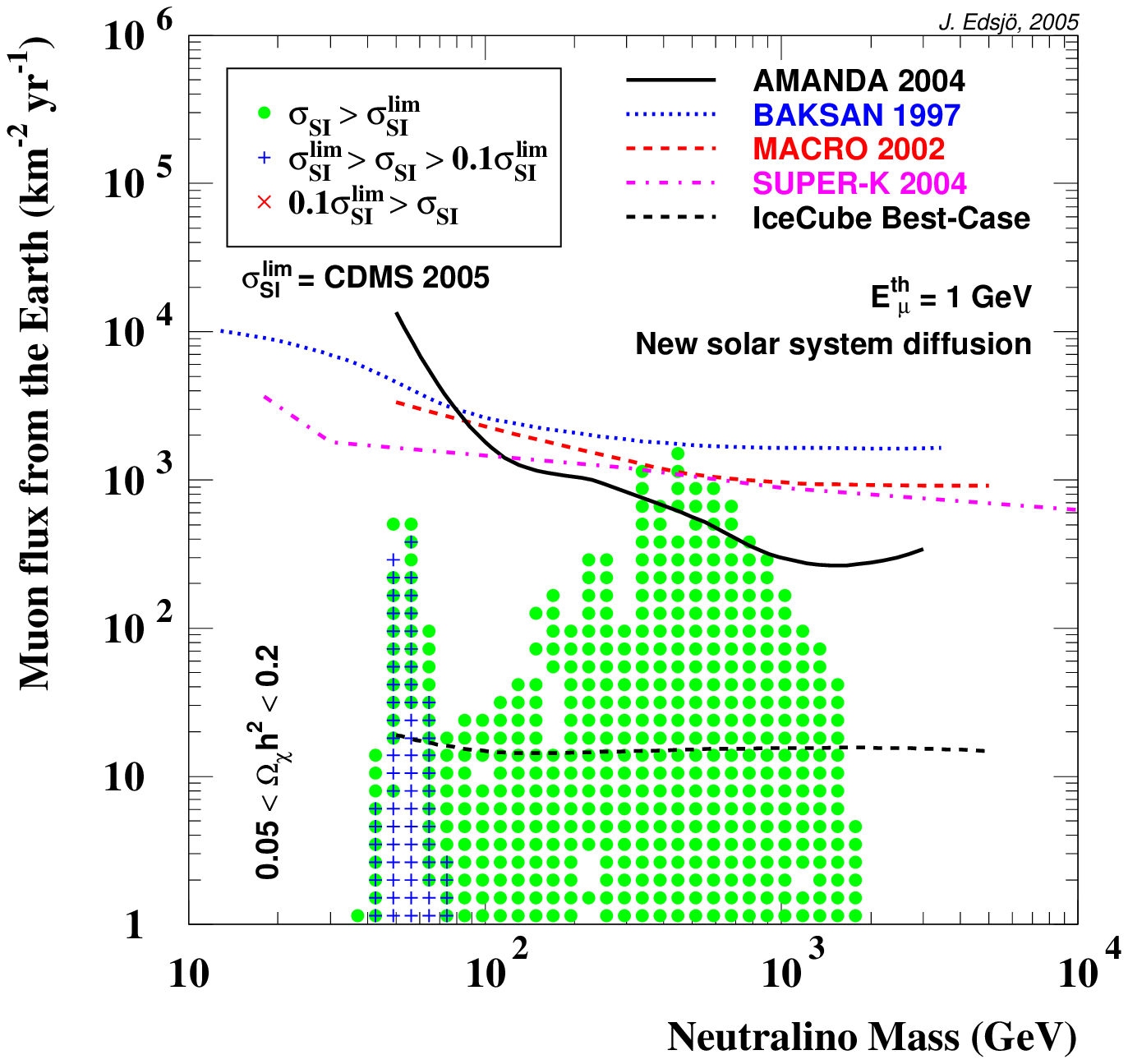}
  \includegraphics[width=.47\textwidth]{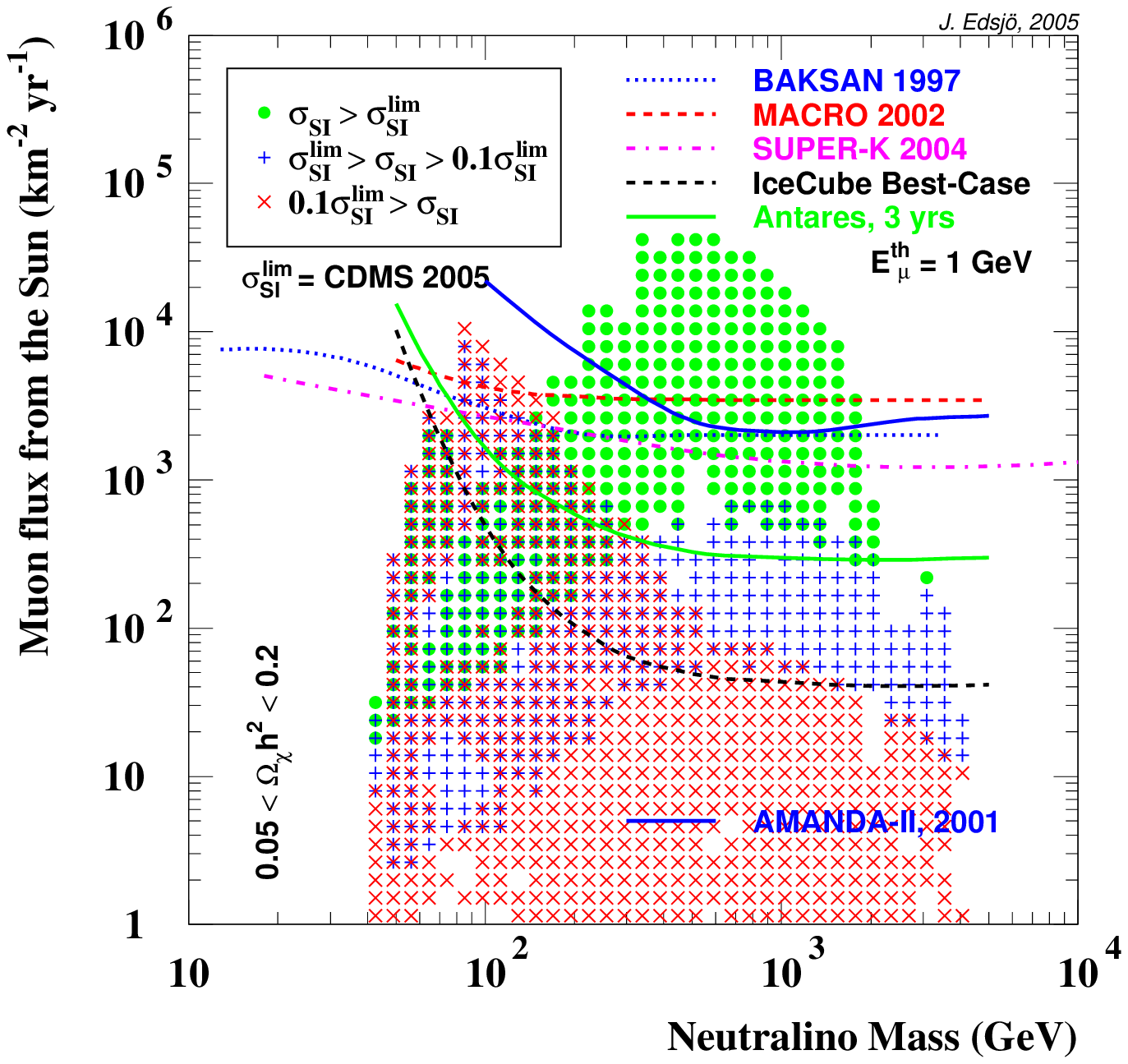}
\caption[]{Limits on the muon flux due to neutrinos from neutralino annihilations in the Earth (left) and the Sun (right) as function of neutralino mass.  The points correspond to predictions from different MSSM models\cite{joakim}.. The dots correspond to models excluded by CDMS-II\cite{cdms} . The plus signs (+) correspond to models testable to a tenfold improved sensitive for direct searches. The crosses (x) are models demanding more than a factor of 10 in increased sensitivity for direct searches.}

\label{fig:wimps}
\end{figure}

There are several questions in particle physics, which can be studied using high energy neutrino telescopes. Perhaps the most important one is the question about the cold dark matter in the universe. The most popular hypothesis for the nature of non-baryonic dark matter is that it consists of the lightest Supersymmetric particle (assuming R-parity conservation), the neutralino, left over from the Big Bang. The mass of the neutralino is expected to be in the GeV-TeV range. The neutralinos scatter weakly on baryonic matter, lose energy and may be trapped in the gravitational field of heavy objects like the Sun and the Earth. The neutralinos accumulated  in the centre of these objects will annihilate and, among other particles, neutrinos will be produced. Observing high energy neutrinos from the Sun or the centre of the Earth could give evidence for  Supersymmetric particles. 
AMANDA analysis of data from 1997-1999 (AMANDA-B10) has resulted in the most sensitive limits on muon flux due to  neutrinos from neutralino annihilation in the centre of the Earth\cite{patrik}. The best corresponding limit for the Sun is given by the Super-Kamiokande  collaboration\cite{superk}. AMANDA has presented a limit  for the Sun based on one year of data (2000)\cite{yulia} and  will,  when all available data are analyzed, be competitive with the best direct search limit by the CDMS-II collaboration\cite{cdms}. Figure ~\ref{fig:wimps} shows the current limits for different MSSM models \cite{joakim}. IceCube  will be able to probe MSSM models not accessible to direct searches. The direct and indirect  methods are complementary since they probe different parts of the neutralino velocity distribution\cite{cdms}.

\begin{figure}[ht]
  \includegraphics[width=.60\textwidth]{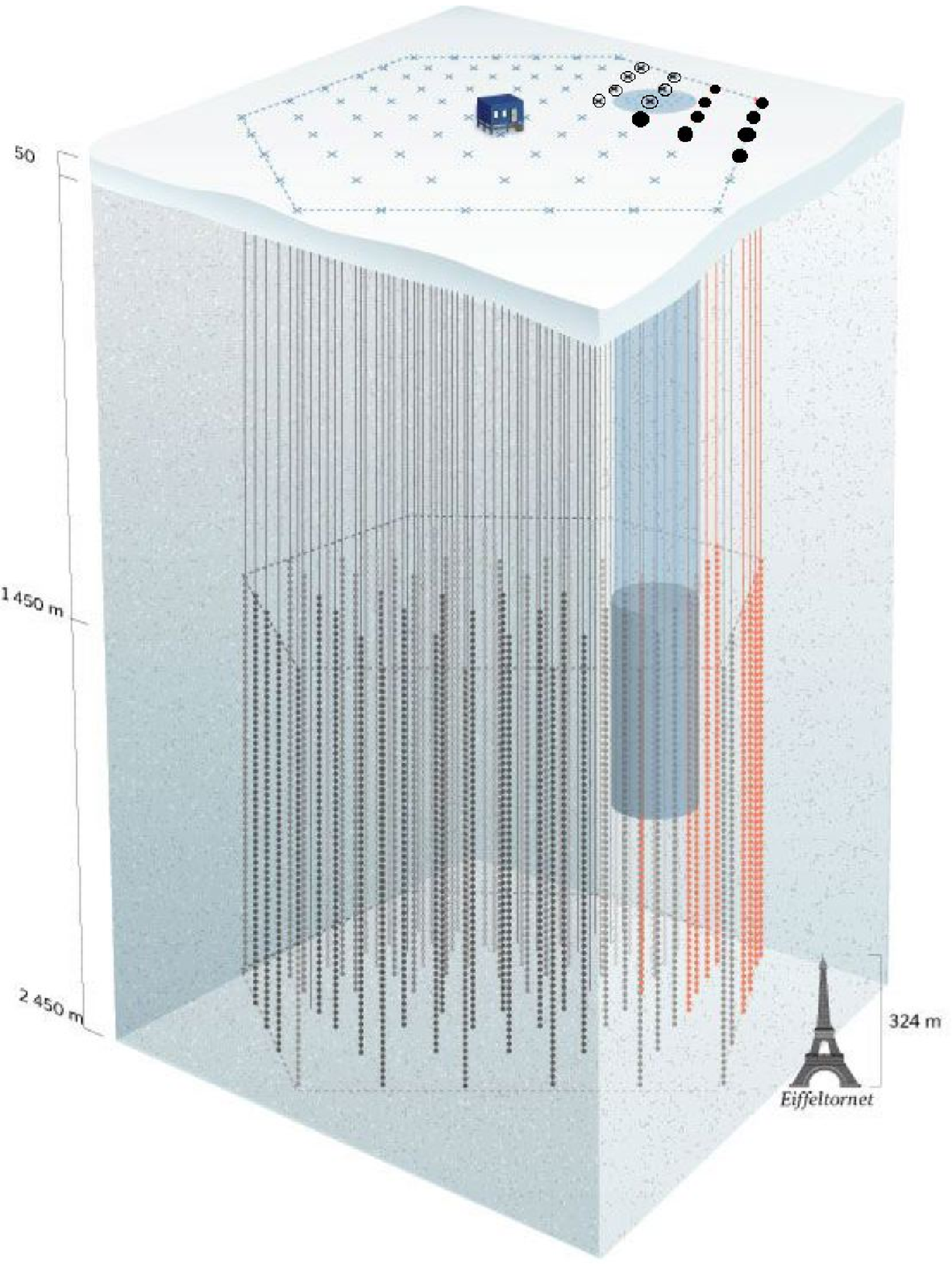}
  \includegraphics[width=.23\textwidth]{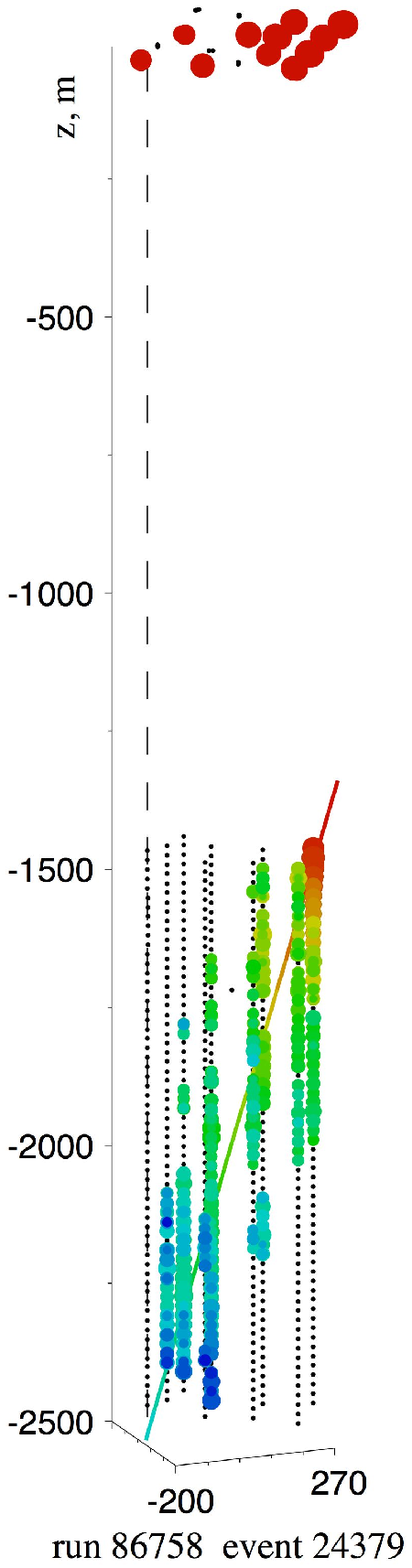} 
\caption{Left: The IceCube Neutrino Observatory with the air shower IceTop at the surface and the In-ice array. The AMANDA telescope is seen inside  IceCube as a darker cylinder. The filled circles on top corresponds to completed strings and IceTop stations (February 2006). The open circles corresponds to additional completed IceTop stations. Right: A large air shower event hitting IceTop modules and the 9-strings.}
\label{fig:icecube}
\end{figure}

\section{IceCube}

The IceCube Neutrino Observatory (figure~\ref{fig:icecube})  at the Amundsen-Scott base at the South Pole, Antarctica,  is the first fully funded  km$^{3}$ sized neutrino detector project.  It  will consist of  the In-ice array deep in the ice and the IceTop air shower array at the surface. In the ice, 80 strings with 60 Digital Optical Modules (DOMs) each will be deployed between depths of 1450 m and 2450 m (17 m between optical modules). The distance between the strings is 125 m. The instrumented deep ice part will cover about one km$^{3}$.  The IceTop air shower array will consist of two ice Cherenkov tanks  placed close to each IceCube hole.  Each tank is instrumented with two of the same DOMs used in the deep ice. The surface array will be used for calibration and background studies as well as for cosmic ray studies using the combined detector.  This is a unique feature of the IceCube Neutrino Observatory. The AMANDA telescope is located inside the volume of IceCube and will be integrated and will extend the reach of the IceCube detector to low energies. Parts of the IceCube detector can be used as veto in order to define a shielded fiducial volume including AMANDA and the inner IceCube strings in order to improve the acceptance for low energy neutrinos from e.g. neutralino annihilation in the Sun. The telescope is modular and new strings will be added into the data acquisition system as soon as they are deployed and commissioned, giving an increasing sensitivity year by year.

The IceCube optical modules are  technically more advanced than  the ones used in AMANDA. They have 25 cm (AMANDA 20 cm) diameter Hamamatsu Photo-multipliers tubes (PMTs) and electronics for digitizing  and time stamping of the (photonic) signals. These Digital Optical Modules (DOMs) transmit the digitized data to the surface via copper cables. The dynamic range exceeds 250 photoelectrons per 10 ns.

\subsection{Construction}
The deployment of the IceCube Neutrino Observatory takes place during the austral summer (November - February).  A new more powerful hot water drill  has been developed with a heat power of 5 MW compared to 2 MW for the AMANDA drill. This new drill melts a 60 cm diameter hole 2500 m in less than 40 h (AMANDA needed  more than 90 h for 2000 m). The hole is water filled from about 50 m below the surface and allows a deployment time for the DOMs of more than 24 h. Water in the hole refreezes from the top down, after about one week near the top and two weeks near the bottom. In the first season (2004/2005) the hot water drill was assembled and tested, and the first IceCube string was successfully deployed (string 21)  together with four IceTop stations, in total 60 + 16 DOMs. During the second deployment season (November 2005 - January 2006) eight new strings were successfully deployed and 12 more IceTop stations completed, giving a total of nine strings and 16 IceTop stations. The configuration of the deployed strings and IceTop stations is shown in figure~\ref{fig:icecube}. Given the experience from the first season the hot water drill had been improved  and the deployment speed raised to about 4 days per string. This is very encouraging  and a deployment rate of 14 or more strings and IceTop stations per year seems within reach. 

\subsection{Experiences from the first deployed string}
All 60 DOMs in the first string (string 21) and the 16 DOMs in the four IceTop stations deployed in February 2005 survived the freezing-in period. The system has been running during 2005 and the performance of the first string is summarized in ref\cite{performans}. The measured  DOM noise rate of about 700 Hz is lower than for the AMANDA modules despite a larger PMT. After imposing an artificial dead time to suppress after pulses, the noise rates are below 400 Hz. This is important for the IceCube supernova trigger which is sensitive to small fluctuations to the general noise rate. The obtained time resolution of  the optical modules is within a few ns over distance of 2.5 km.

The deployed modules performs as well as, or even better than, expected.

 \begin{figure}
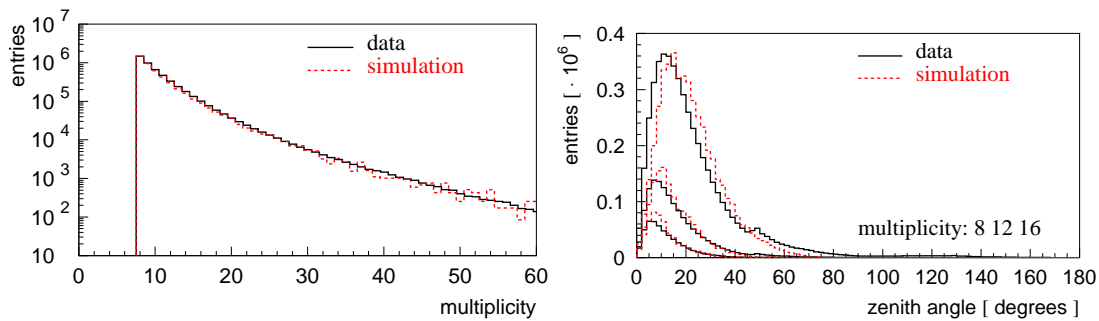

  \includegraphics[width=.47\textwidth]{nch.epsi}
  \includegraphics[width=.47\textwidth]{data.epsi} 

\caption{(left) The number of hit DOMs in muon tracks, compared with
simulation for the first IceCube string (21).  At least 8 hits are required for a trigger. (right) The
zenith angle distribution for muons events with multiplicity at least
8, 12, 16 or 20 hits, compared with the corresponding simulations.
Here 180$^0$ points straight downward.}
\label{fig:muons}
\end{figure}

Muons passing string 21 were reconstructed and compared to the expectation from simulated events.  In figure~\ref{fig:muons} (left) the DOM multiplicity for muon tracks in  simulation and experimental data is compared and a reasonable agreement can be seen. The zenith angle distribution is shown in figure~\ref{fig:muons} (right ) for different multiplicity cuts (azimuth is not possible to reconstruct with only one string). The simulation was done with AMANDA software and will be improved over time probably resulting in an even better agreement.

The string 21 is about 328 m from the centre of AMANDA and off-line analysis of events in coincidence in string 21 and AMANDA has been done.

With only one single string the amount of physics analysis was limited, however, in the data taken with string 21 during 2005 two upward going tracks with high multiplies  (35 and 50 out of 60) were found, compatible with the expected flux of  atmospheric neutrinos (preliminary).

\subsection{The nine string IceCube}
The new eight strings are now in an intense  period of verification. The volume instrumented by the nine strings is already now the largest in the world.  The 540 DOMs in ice and 64 DOMs in IceTop tanks together with the 677 optical modules in AMANDA constitute a very interesting scientific instrument. A recorded air shower hitting the Icetop stations and the in ice strings is shown in figure~\ref{fig:icecube}.

\section{Conclusions}
The AMANDA neutrino telescope is taking data for the seventh year. The experience from one year of operating the first IceCube string and the four IceTop stations is very good and shows that the equipment performs as expected. The now deployed nine strings and IceTop stations will allow a more detailed study of the system as well as taking real data for physics analysis. The construction of the IceCube neutrino observatory is proceeding very well.

\section{Acknowledgements}

 This research was supported by the Deutsche Forschungsgemeinschaft
(DFG); German Ministry for Education and Research; Knut and Alice
Wallenberg Foundation, Sweden; Swedish Research Council; Fund for Scientific Research
(FNRS-FWO), Flanders Institute (IWT), Belgian Federal Office for
Scientific, Technical and Cultural affairs (OSTC), Belgium. UC-Irvine
AENEAS Supercomputer Facility; University of Wisconsin Alumni Research
Foundation; U.S. National Science Foundation, Office of Polar
Programs; U.S. National Science Foundation, Physics Division;
U.S. Department of Energy.

\end{document}